\documentclass[10pt,conference]{NEMO}
\usepackage{graphicx}
\usepackage{multirow}
\usepackage{cite}
\usepackage{bm}
\usepackage{amsmath,amssymb,amsfonts}
\usepackage{algorithmic}
\usepackage{graphicx}
\usepackage{textcomp}
\usepackage{xcolor}

\ifCLASSINFOpdf
\else
\fi

\makeatletter
\def\ps@IEEEtitlepagestyle{%
\def\@oddfoot{\mycopyrightnotice}%
\def\@evenfoot{}%
}
\def\mycopyrightnotice{%
{\footnotesize 979-8-3503-5183-5/24/\$31.00~\copyright~2024 IEEE \hfill} %
\gdef\mycopyrightnotice{}
}
\begin{document}
%
\title{Capacity Based Design of Slot Array Antennas}

\author{Volodymyr~Shyianov,~Bamelak~Tadele,~Vladimir~I.~Okhmatovski,~Faouzi~Bellili,~and~Amine~Mezghani\\
		\\
University~of~Manitoba,~Winnipeg,~Manitoba,~R3T~5V6,~Canada}




\maketitle

\begin{abstract}
Historically, the design of antenna arrays has evolved separately from Shannon theory. Shannon theory adopts a probabilistic approach in the design of communication systems, while antenna design approaches have relied on deterministic Maxwell theory alone. In this paper, we introduce a new approach to the design of antenna arrays based on information theoretic metrics. To this end, we develop a statistical model suitable for the numerical optimization of antenna systems. The model is utilized to obtain the signal-to-noise ratio (SNR), find the optimal power allocation scheme, and establish the associated Shannon capacity. We demonstrate the utility of the new approach on a connected array of slot antennas. To find the impedance matrix of the slot array, we further develop a fast numerical technique based on the analytical form of the spectrum of magnetic current. The utilized spectral approach, albeit its simplicity, shows good match compared with full wave electromagnetic simulation. 
\end{abstract}


%
\IEEEpeerreviewmaketitle

\section{Introduction}
 Consider a transmitter equipped with $N_t$ antenna elements which aims to communicate with a receiver that is endowed with $N_r$ antenna elements. The mathematical representation of the underlying MIMO channel is given by:
\begin{equation}\label{eq:MIMO_channel}
    \mathbf{y} = \mathbf{H}\mathbf{x}+\mathbf{n},
\end{equation}
where $\mathbf{n}$ represents the additive noise. This widely adopted noisy linear model has lead to great developments both from signal processing and information theory perspectives \cite{tse2005fundamentals,heath2018foundations}. While this model captures the essence of Maxwell's equations, the precise description of the channel matrix $\mathbf{H}$ and the statistics of the noise $\mathbf{n}$ remains an active research topic. Multi-port communication theory, first introduced in \cite{wallace2004mutual} and popularized by \cite{ivrlavc2010toward}, offers a consistent approach to incorporate the physics of radio-communication into the model of the channel matrix and the noise statistics. In this model, we have three interfaces between the transmitter and the receiver. The first and third interfaces contain the multi-port networks that aim to optimize (through different criteria) the link between the transmit/receive signals and their respective antennas. The middle interface is a multi-port network that incorporates the physics of propagation as well as the parameters of the antennas in use. Together the communication channel is given by the relationship between the generator signal (voltage/current) and the load signal. This multi-port channel representation has lead to new insights in beamforming \cite{ivrlavc2010toward}, revealed the impact of the antenna size \cite{shyianov2021achievable,akrout2023super} on the achievable data rate, and was used to improve algorithms for channel estimation \cite{tadele2023channel}. Further, by merging multi-port communication theory with information theory, the achievable rate criteria was used for the design of the optimal matching networks in SISO systems \cite{taluja2010information,shyianov2021achievable} as well as the analysis of coupling in wideband SIMO systems
\cite{saab2019capacity}. In this paper, we design an array of slot antennas by optimizing the shannon capacity using multi-port communication theory. Slot array antennas have previously been investigated in the context of connected arrays \cite{cavallo2011connected}. Such arrays have demonstrated large, theoretically infinite, bandwidth with the increase of the array size. As such, the connected slot structure is a promising design for massive MIMO base station arrays.   

\section{System Model}
Based on the model (\ref{eq:MIMO_channel}) we can use multiport communication theory \cite{ivrlavc2010toward} to represent the input signal, $\mathbf{x}$, by transmit voltage sources, and the output signal, $\mathbf{y}$, by the voltages received at the load. Under this paradigm, the channel matrix $\mathbf{H}$, at any frequency, can be found from cascaded multiport networks and is given by:
\begin{equation}\label{eq:Impedance_MIMO_channel}
    \mathbf{H}(f) ~=~ \mathbf{Z}_{\mathrm{L}}(\mathbf{Z}_{\mathrm{R}}+\mathbf{Z}_\mathrm{L})^{-1}\mathbf{Z}_{\mathrm{RT}}(\mathbf{Z}_{\mathrm{T}}+\mathbf{Z}_{\mathrm{S}})^{-1}. 
\end{equation}
Here the matrices $\mathbf{Z}_{\mathrm{S}}(f)$ and $\mathbf{Z}_{\mathrm{L}}(f)$ represent the source and load multiport networks, $\mathbf{Z}_{\mathrm{T}}(f)$ and $\mathbf{Z}_{\mathrm{R}}(f)$ represent the impedance matrices of the transmit and receive arrays, and $\mathbf{Z}_{\mathrm{RT}}(f)$ is a multiport network for the medium of propagation. Assuming a rich-scattering environment the matrix $\mathbf{Z}_{\mathrm{RT}}(f)$ is given by:
\begin{equation}\label{eq:FF-mutual-impedance-gaussian}
\mathbf{Z}_{\mathrm{RT}}(f) = ~\frac{c}{2{\pi}fd^{\frac{\alpha}{2}}}~\Re{\{\mathbf{Z_{\mathrm{R}}}(f)\}}^{\frac{1}{2}}~\mathbf{F}~\Re{\{\mathbf{Z_{\mathrm{T}}}(f)\}}^{\frac{1}{2}}~~[\Omega].
\end{equation}
where $\mathbf{F}_{ij}\sim\mathcal{CN}(0,1)$ and are statistically independent of one another. Furthermore, the noise vector $\mathbf{n}$ from (\ref{eq:MIMO_channel}) is given by:
\begin{equation}\label{Impedance_MIMO_Noise}
    \mathbf{n}(f) ~=~ \mathbf{v}_\mathrm{IN}~+~\mathbf{Z}_{\mathrm{L}}(\mathbf{Z}_{\mathrm{R}}+\mathbf{Z}_{\mathrm{L}})^{-1}\mathbf{v}_\mathrm{EN},
\end{equation}
where $\mathbf{v}_\mathrm{IN}$ and $\mathbf{v}_\mathrm{EN}$ are the intrinsic and extrinsic noise sources. The corresponding noise correlation matrix is given by:
\begin{equation}\label{eq:Impedance_noise_correlation}
\begin{aligned}
\mathbf{R}_{\mathbf{n}}(f) 
& = 4k_bT\mathbf{Z}_{\mathrm{L}}(\mathbf{Z}_{\mathrm{R}}+\mathbf{Z}_{\mathrm{L}})^{-1}\Re\{\mathbf{Z}_{\mathrm{R}}(f)\}(\mathbf{Z}_{\mathrm{R}}+\mathbf{Z}_{\mathrm{L}})^{-\mathsf{H}}\mathbf{Z}_{\mathrm{L}}^{\mathsf{H}} \\
&+ 4k_{\textrm{b}}T(N_f - 1)R_{in}\mathbf{I}.
\end{aligned}
\end{equation}

\section{Connected Slot Array}
\begin{figure}[h!]
    \centering
    \includegraphics[scale=0.5]{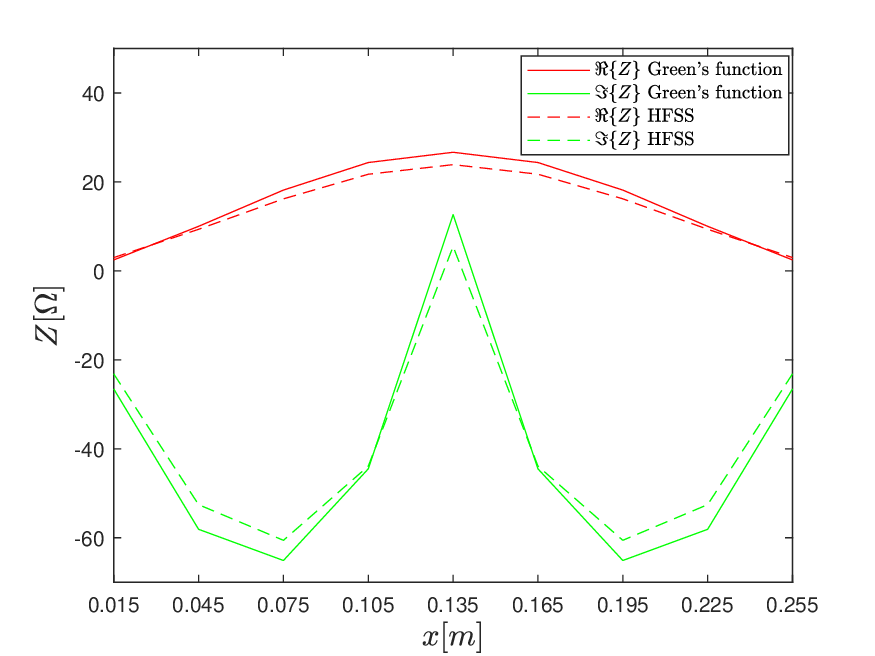}
    \includegraphics[scale=0.5]{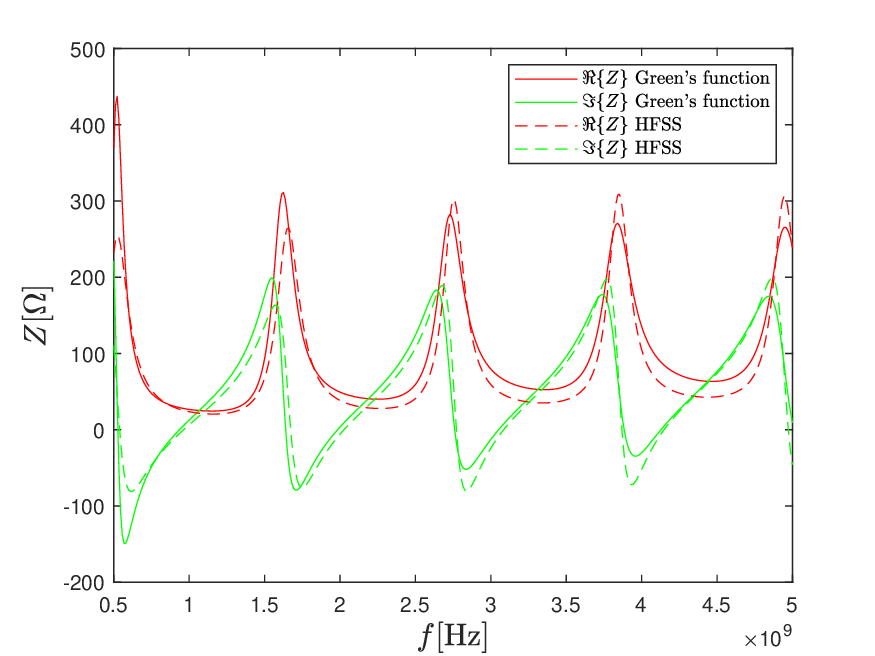}
    \caption{Imdedance of the connected slot array in free space as a function of position and frequency compared with HFSS simulation}
    \label{fig:Slot Hfss}
    \vspace{-0.4cm}
\end{figure}

Consider an x-oriented slot in a perfectly conducting (PEC) plate.
We denote the uniform cross-section width of the slot by $w_s$ and assume that it is much smaller than wavelength. The structure is excited by the $y$-directed electric current of length $w_s$, placed at the origin, $\mathbf{J} = I_0\delta(x){rect}_{w_s}(y)\hat{y}$. To derive the magnetic current distribution on the finite structure we proceed to first determine the magnetic current of an infinite slot in the x-direction, embedded in an infinite ground plane.
Making use of the equivalence theorem, the $(x,y)$ plane can be replaced by the equivalent distribution of surface currents.
By filling the slot region with the PEC we are left only with magnetic currents $m(x,y)\hat{x}$ in the top half-space and $-m(x,y)\hat{x}$ in the bottom half-space, separated by an infinite PEC plate\footnote{Here we assumed currents are perfectly polarized along $\hat{x}$ which is well justified when $w_s$ is much smaller than wavelength}. By enforcing the continuity of the total magnetic field at $(y = 0)$, we have the following integral equation,
\begin{equation}\label{eq:integral equation space}
    \int_{-\infty}^{\infty} \int_{-w_s / 2}^{w_s / 2} g_{x x}\left(x-x^{\prime},-y^{\prime}\right) m\left(x^{\prime}, y^{\prime}\right) d x^{\prime} d y^{\prime} = I_0\delta(x).
\end{equation}
Now, the expression for green's function at the slot $g_{x x}(x,y)$ can be derived by using the image principle and again relying on the continuity of the magnetic field,
\begin{equation}
    g_{x x}(x,y) = 4g_{x x}^{\textrm{FS}}(x,y)
\end{equation}
where $g_{x x}^{\textrm{FS}}(x,y)$ is the homogeneous space green's function. The magnetic currents are further assumed to have separable space dependence, 
\begin{equation}
    m(x,y) = v(x)m_t(y),
\end{equation}
where $m_t(y)$ is chosen to satisfy the quasi-static edge singularity, with normalization chosen such that $v(x)$ represents the total voltage drop across any two points along the slot,
\begin{equation}
    m_t(y)=-\frac{2}{w_s \pi} \frac{1}{\sqrt{1-\left(\frac{2 y}{w_s}\right)^2}}.
\end{equation}
Taking the Fourier transform of the integral equation in (\ref{eq:integral equation space}) we obtain \cite{neto2003green},
\begin{equation}
    V(k_x)D(k_x) = I_0,
\end{equation}
where $V(k_x)$ is the spatial Fourier Transform of of the voltage $v(x)$ and 
\begin{equation}\label{eq: inverse voltage spectrum}
    \begin{aligned}
D\left(k_x\right)=\frac{1}{k_0 \zeta_0} & \left(k_0^2-k_x^2\right) J_0\left(\frac{w_s}{4} \sqrt{k_0^2-k_x^2}\right) \\
& \times H_0^{(2)}\left(\frac{w_s}{4} \sqrt{k_0^2-k_x^2}\right).
\end{aligned}
\end{equation}
The voltage along the slot can be calculated by use of the inverse Fourier Transform,
\begin{equation}\label{eq: voltage free space}
    v(x)=\frac{I_0}{2 \pi} \int_{-\infty}^{\infty} \frac{e^{-j k_x x}}{D\left(k_x\right)} d k_x.
\end{equation}
To numerically evaluate the expression in (\ref{eq: voltage free space}) we need to avoid the poles at $k_x=\pm k_0$, and require Cauchy theorem to deform the original integration path into a convergent path in the complex plane.\footnote{It is convenient to choose a contour to be close to the real line to improve stability.} 
The mutual impedance between any two antenna elements $k$ and $m$ within the array is then easily computed as,
\begin{equation}
    (\mathbf{Z}_{\textrm{A}})_{k,m} = \frac{v((k-m)d_x)}{I_0},
\end{equation}
here $\mathbf{Z}_{\textrm{A}}$ refers to either the transmitter or the receiver.
 By adding two extra ports at $x = 0$ and $x = L$ we can enforce the boundary condition at the edges through short circuiting the two extra ports. The impedance matrix of the slot of finite length $L$ can be determined as,
\begin{equation}\label{eq:finite admittance}
    \mathbf{Z}_{\textrm{A}}^{f} = \mathbf{Z}_{\mathrm{L}}(\mathbf{Z}_{\mathrm{A}}+\mathbf{Z}_\mathrm{L})^{-1}\mathbf{Z}_{\textrm{A}}\mathbf{I},
\end{equation}
where $\mathbf{Z}_{\textrm{L}}$ is diagonal with zeros for the first and last extra port at $x = 0$ and $x = L$ (short circuit termination) and infinity otherwise. Practically, open circuit termination is achieved by taking sufficiently large resistive termination ($R\approx 10^5$). A different approach, that does not require additional parameters is to first compute the admittance matrix of the array,
\begin{equation}\label{eq:admittance}
    \mathbf{Y}_{\textrm{A}} = \mathbf{Z}_{\textrm{A}}^{-1}
\end{equation}
and then remove the first and last row and column corresponding to $x = 0$ and $x = L$ before inverting back to get the impedance matrix. Figure.~\ref{fig:Slot Hfss}, show a close match between analytical formulas in  (\ref{eq:admittance}) and (\ref{eq:finite admittance}) and the impedance simulated through HFSS.
\noindent

\section{Capacity Analysis}
The Shannon capacity is given by \cite{heath2018foundations},\cite{gallager1968information}:
\begin{equation}\label{AWGN_Capacity-noise-correlation}
    C = \sum_{l=0}^{L-1}\int_{-\mathrm{W}/2}^{\mathrm{W}/2}{\log_2\Big(1+ \mathrm{SNR}_{l}(f)\Big)\textrm{d}f}\,\,[{\textrm{bits/s}}],
\end{equation}
where 
\begin{equation}
    \textrm{SNR}_{l}(f) = P_{t,l}(f)\,\nu_l\Big[\mathbf{H}^{\mathsf{H}}(f)\,\mathbf{R}_{\mathbf{n}}^{-1}(f)\,\mathbf{H}(f)\Big],
\end{equation}
represents the SNR of the $l^{th}$  effective sub-channel, $L$ is the number of non-zero eigenvalues and $\mathrm{W}$ is the signal bandwidth. The operator $\nu_l[.]$ returns the $l^{th}$ ordered eigenvalue of a matrix. The power allocation $P_{t,l}(f)$, for $l=0, \ldots, L-1,$ is determined from the waterfilling solution:
\begin{equation}
    P_{t,l}(f) = \left(0,\,{\frac{1}{B}} - {\frac{1}{\nu_{l}[\mathbf{H}^{\mathsf{H}}(f)\,\mathbf{R}^{-1}_{\textbf{n}}(f)\,\mathbf{H}(f)]}} \right) 
\end{equation}
and $B$ is chosen to satisfy the total power constraint:
\begin{equation}
\int_{-\mathrm{W}/2}^{\mathrm{W}/2}\sum_{l=0}^{L-1}\,P_{t,l}(f)\textrm{d}f = P.
\end{equation}

\section{Simulation Results and Discussion}
\begin{figure}[h!]
    \centering
    \includegraphics[scale=0.5]{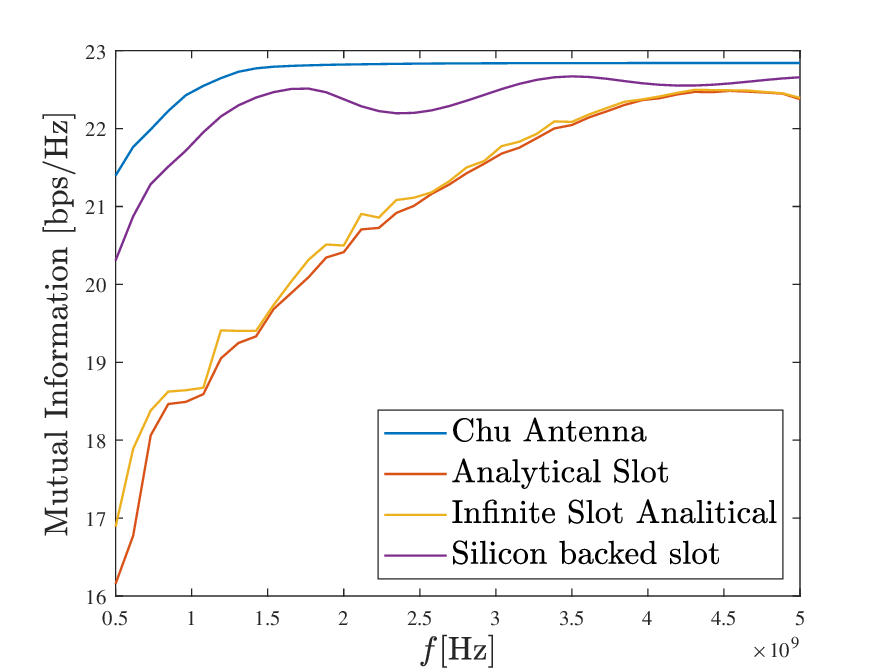}
    \caption{Spectral efficiency vs frequency for the connected slot array in free-space as well as backed by a silicon half-space compared to the ultrawideband array of Chu's antnnas }
   \label{fig:Chu vs slot}
    \vspace{-0.65cm}
\end{figure}

 In the simulation, we consider broadband transmit/receive connected antenna arrays operating between $f_{\textrm{min}}=500\,[\textrm{MHz}]$ and $f_{\textrm{max}}=5\,[\textrm{GHz}]$ which corresponds to a decade of bandwidth. We also set the total power to $P_{\textrm{max}} = 2\,[\textrm{W}]$, the path-loss exponent $\alpha$ to 3.5, and the separation distance between the transmit and receive arrays to $d=30\,\lambda_{\textrm{min}}$. For simpler interpretation, we assume the large-scale path-loss in not frequency dependant (we remove the frequency from (\ref{eq:FF-mutual-impedance-gaussian})). In Fig.\ref{fig:Chu vs slot}, we plot the spectral efficiency of the MISO system over the desired band of frequencies for three different types of slot antennas and compare their performance with the Chu's antnna array. The types of antennas are slot antenna array in free space, slot array backed by silicon half-space, which corresponds to $\epsilon_r = 11.7$, and theoretical impedance matrix derived from the infinite slot. All antenna arrays have $N = 64$ elements, which brings the total array size to $1.92[\textrm{m}]$. The length was chosen on purpose to make the element spacing within the array be $\lambda_{\textrm{max}}/2$, where $\lambda_{\textrm{max}}$ is the wavelength at $f = 5[\textrm{GHz}]$,the largest frequency in the band. We also assume a single receive antenna matched to the LNA. At first, its interesting to point out that the mutual information of the theoretical infinite slot array and finite array become very close as the frequency increases which is intuitively expected as the array is electrically large at high frequency, approaching the theoretical limit of an infinite slot. What is more interesting is the improved performance brought by silicon half-space as opposed to simple free-space array. This observation can be attributed to the fact that silicon half-space is acting as a MN network, increasing the overall bandwidth of the array.
\section{Conclusion}
In this paper, we introduced a new approach to the design of antenna arrays based on information theoretic metrics. We developed a statistical model suitable for numerical optimization of antenna systems. The model is utilized to obtain the signal-to-noise ratio (SNR), find the optimal power allocation, and establish the associated Shannon capacity. We demonstrate the utility of the new approach on a connected array of slot antennas. For the determination of the impedance matrix of the slot array, we developed a fast numerical approach based on the analytical form of the magnetic current.

\bibliographystyle{IEEEtran}
\bibliography{IEEEabrv,references}

\begin{thebibliography}{10}
\providecommand{\url}[1]{#1}
\csname url@samestyle\endcsname
\providecommand{\newblock}{\relax}
\providecommand{\bibinfo}[2]{#2}
\providecommand{\BIBentrySTDinterwordspacing}{\spaceskip=0pt\relax}
\providecommand{\BIBentryALTinterwordstretchfactor}{4}
\providecommand{\BIBentryALTinterwordspacing}{\spaceskip=\fontdimen2\font plus
\BIBentryALTinterwordstretchfactor\fontdimen3\font minus
  \fontdimen4\font\relax}
\providecommand{\BIBforeignlanguage}[2]{{%
\expandafter\ifx\csname l@#1\endcsname\relax
\typeout{** WARNING: IEEEtran.bst: No hyphenation pattern has been}%
\typeout{** loaded for the language `#1'. Using the pattern for}%
\typeout{** the default language instead.}%
\else
\language=\csname l@#1\endcsname
\fi
#2}}
\providecommand{\BIBdecl}{\relax}
\BIBdecl

\bibitem{tse2005fundamentals}
D.~Tse and P.~Viswanath, \emph{Fundamentals of wireless communication}.\hskip
  1em plus 0.5em minus 0.4em\relax Cambridge university press, 2005.

\bibitem{heath2018foundations}
R.~W. Heath~Jr and A.~Lozano, \emph{Foundations of MIMO communication}.\hskip
  1em plus 0.5em minus 0.4em\relax Cambridge University Press, 2018.

\bibitem{wallace2004mutual}
J.~W. Wallace and M.~A. Jensen, ``Mutual coupling in mimo wireless systems: A
  rigorous network theory analysis,'' \emph{IEEE transactions on wireless
  communications}, vol.~3, no.~4, pp. 1317--1325, 2004.

\bibitem{ivrlavc2010toward}
M.~T. Ivrla{\v{c}} and J.~A. Nossek, ``Toward a circuit theory of
  communication,'' \emph{IEEE Transactions on Circuits and Systems I: Regular
  Papers}, vol.~57, no.~7, pp. 1663--1683, 2010.

\bibitem{shyianov2021achievable}
V.~Shyianov, M.~Akrout, F.~Bellili, A.~Mezghani, and R.~W. Heath, ``Achievable
  rate with antenna size constraint: Shannon meets chu and bode,'' \emph{IEEE
  Transactions on Communications}, 2021.

\bibitem{akrout2023super}
M.~Akrout, V.~Shyianov, F.~Bellili, A.~Mezghani, and R.~W. Heath,
  ``Super-wideband massive mimo,'' \emph{IEEE Journal on Selected Areas in
  Communications}, 2023.

\bibitem{tadele2023channel}
B.~Tadele, V.~Shyianov, F.~Bellili, and A.~Mezghani, ``Channel estimation with
  tightly-coupled antenna arrays,'' in \emph{ICASSP 2023-2023 IEEE
  International Conference on Acoustics, Speech and Signal Processing
  (ICASSP)}.\hskip 1em plus 0.5em minus 0.4em\relax IEEE, 2023, pp. 1--5.

\bibitem{taluja2010information}
P.~S. Taluja and B.~L. Hughes, ``Information theoretic optimal broadband
  matching for communication systems,'' in \emph{2010 IEEE Global
  Telecommunications Conference GLOBECOM 2010}.\hskip 1em plus 0.5em minus
  0.4em\relax IEEE, 2010, pp. 1--6.

\bibitem{saab2019capacity}
S.~Saab, A.~Mezghani, and R.~W. Heath, ``Capacity based analysis of a wideband
  simo system in the presence of mutual coupling,'' in \emph{2019 IEEE Global
  Communications Conference (GLOBECOM)}.\hskip 1em plus 0.5em minus 0.4em\relax
  IEEE, 2019, pp. 1--6.

\bibitem{cavallo2011connected}
D.~Cavallo, ``Connected array antennas: Analysis and design,'' 2011.

\bibitem{neto2003green}
A.~Neto and S.~Maci, ``Green's function for an infinite slot printed between
  two homogeneous dielectrics. i. magnetic currents,'' \emph{IEEE Transactions
  on Antennas and Propagation}, vol.~51, no.~7, pp. 1572--1581, 2003.

\bibitem{gallager1968information}
R.~G. Gallager, \emph{Information theory and reliable communication}.\hskip 1em
  plus 0.5em minus 0.4em\relax Springer, 1968, vol. 588.

\end{thebibliography}

\end{document}